\documentclass[aps,prd,amsfonts,preprint,
eqsecnum,nofootinbib]
{revtex4}
\usepackage{graphics,epsfig}
\begin{document}
\preprint{WM-02-108}
%
\title{\vspace*{0.3in}
Supersymmetric Noncommutative QED and Lorentz Violation\vskip
0.1in}
\author{Carl E. Carlson}
\email[]{carlson@physics.wm.edu}
\author{Christopher D. Carone}
\email[]{carone@physics.wm.edu}
\affiliation{Nuclear and Particle Theory Group, Department of
Physics, College of William and Mary, Williamsburg, VA 23187-8795}
\author{Richard F. Lebed}
\email[]{Richard.Lebed@asu.edu}
\affiliation{Department of Physics and Astronomy, Arizona State 
University, Tempe, AZ 85287-1504}

\date{September 6, 2002}
\begin{abstract}
We consider Lorentz-violating operators induced at the loop level in
softly-broken supersymmetric noncommutative QED.  Dangerous operators
forbidden in the supersymmetric limit are generated via finite
corrections, with the scale of supersymmetry breaking serving as a
gauge-invariant regulator.  We compare the most dangerous loop effects
to those obtained in noncommutative theories truncated by a
momentum-space cutoff, and find significantly improved bounds. 
\end{abstract}
\pacs{}
\maketitle


\section{Introduction}\label{sec:intro}


The idea that spacetime may be modified in nontrivial ways at 
distance scales that are accessible at high energy colliders has 
led to recent interest in the phenomenology of noncommutative 
field theories~\cite{ncqed1,ncsm,CHKLT,MPR,ABDG,CCL,ncxd,GMW,posp2}.  
Such theories are known to arise in string theory~\cite{SW} and to have 
interesting properties~\cite{dn}.  The phenomenology of these 
theories is determined by a real, antisymmetric matrix $\theta$ 
which defines the fundamental spacetime commutation relation:
\begin{equation}
[\hat{x}^\mu \, , \, \hat{x}^\nu] = i \, \theta^{\mu\nu} \,\,\, .
\label{eq:canonical}
\end{equation}
Notice that the coordinate $x^\mu$ has been promoted to an operator 
$\hat{x}^\mu$.   Field theories on a noncommutative space can be
constructed in terms of fields that are functions of commuting spacetime
coordinates provided that ordinary multiplication is promoted to star 
multiplication. The star product corresponding to Eq.~(\ref{eq:canonical}) 
is given by
\begin{equation}
(f \star g)(x) = f(x)\, \exp\left[\frac{i}{2} \stackrel{\leftarrow}
{\partial_\mu} \theta^{\mu\nu} \stackrel{\rightarrow}{\partial_\nu}\right]\,
g(x) \,\,\, ,
\label{eq:moyal}
\end{equation}
for any two functions $f$ and $g$, the well-known Moyal-Weyl result.  A 
field theory action is then of the form 
\begin{equation}
S = \int d^4 x \, {\cal L}(\phi(x), \partial_\mu \phi(x))_\star \,\,\, ,
\label{eq:acform}
\end{equation}
where the $\star$ subscript indicates that all multiplications between
fields are defined as in Eq.~(\ref{eq:moyal}).  The specific form
of ${\cal L}$ is fixed by the usual requirements of invariance under
the local and global symmetries of the theory.

By far, the largest number of studies in noncommutative phenomenology have
been directed toward noncommutative QED (NCQED), both in 
four~\cite{ncqed1} or more~\cite{ncxd} dimensions.  The field 
strength tensor in NCQED is given by
\begin{equation}
F_{\mu\nu}=\partial_\mu A_\nu - \partial_\nu A_\mu - 
i e [A_\mu\stackrel{\star}{\,,}A_\nu] \, ,
\label{eq:fmunu}
\end{equation}
indicating the existence of three- and four-photon vertices.  The
observable effects of these new interactions have been considered in
collider studies by a number of authors~\cite{ncqed1}.

On the other hand, low-energy non-accelerator experiments may provide
much more stringent~\cite{CHKLT}, if not
insurmountable~\cite{MPR,ABDG,CCL}, bounds on the size of $\theta$.
The most notable phenomenological feature of canonical noncommutative
field theories is the violation of Lorentz invariance following from
Eq.~(\ref{eq:canonical}).  Both $\theta^{i0}$ and $\epsilon^{ijk}
\theta_{jk}$ are fixed three-vectors that define preferred directions
in a given Lorentz frame. Anisimov, Banks, Dine and
Graesser~\cite{ABDG} have pointed out that operators of lower mass
dimension are generated via loop effects in noncommutative theories
and these operators are severely constrained by low-energy searches
for the violation of Lorentz invariance~\cite{CK}.  If the loop
integrals are evaluated without a momentum space cutoff, the most
dangerous Lorentz-violating operators receive contributions that are
independent of the scale of $\theta$.  This is due to powers of
$\theta^{-1}$ that appear after integration due to the peculiar
momentum dependence of the noncommutative vertices.  Experimental
bounds in this case cannot be evaded by raising the scale of
noncommutativity, and the underlying theory is excluded~\cite{ABDG}.
On the other hand, if a momentum-space cutoff $\Lambda$ is used, for
example, to take into account a change in the physics at a low Planck
scale, then the Lorentz-violating effects depend on the scale of
noncommutativity through the product $\theta \Lambda^2$, which can be
bounded.  In NCQED, consideration of the most dangerous
operator
\begin{equation}
{\cal O}_1 = m_e \theta^{\mu\nu} \bar\psi \sigma_{\mu\nu}\psi \,\,\, ,
\label{eq:danger}
\end{equation}
where $\psi$ represents the electron field and $m_e$ the electron mass,
leads to the bound $\theta\Lambda^2 < 10^{-19}$~\cite{ABDG}; even tighter 
bounds have been shown to arise in noncommutative QCD~\cite{CCL}.  These 
results suggest that if Lorentz-violating noncommutativity is realized in 
nature, the size of $\theta$ is much smaller than one would expect 
from naive dimensional arguments.

It is natural to question the reliability of bounds obtained via regulating
a gauge theory with a hard, ultraviolet cutoff.  Such a cutoff violates the 
gauge invariance of the theory, and is not defined precisely in terms of 
physical quantities.  It is the purpose of this Letter to investigate whether 
the phenomenological conclusions described above are altered substantially when
one employs a cutoff that is both physical and preserves 
gauge invariance. 

We therefore focus on softly-broken supersymmetric NCQED.  In the
supersymmetric limit~\cite{ncs}, one may show that the most dangerous
Lorentz-violating operator in Eq.~(\ref{eq:danger}) is forbidden:
there is no way to construct an $F$ or $D$ term using
$\theta^{\mu\nu}$, superfields, and derivatives, that reduces to the
desired Dirac and Lorentz structure. When supersymmetry is softly
broken by giving the superpartners a common mass $M$, the dangerous
operator in Eq.~(\ref{eq:danger}) is again generated; however,
supersymmetric cancellations eliminate contributions from the
ultraviolet part of the loop integrals.  Thus, $M$ serves as an
effective cutoff that preserves the gauge invariance of the theory. In
the next section we adopt this framework in computing the operator in
Eq.~(\ref{eq:danger}), which is generated at the two-loop level.  The
dependence on $\theta M^2$ differs from what one would expect given a
hard cutoff, and leads to a stronger bound on the size of $\theta$.
We then comment upon Lorentz-violating corrections to the photon
propagator and summarize our conclusions.

\section{A Dangerous Operator}

\begin{figure}[ht]

\epsfxsize 3.35 in \epsfbox{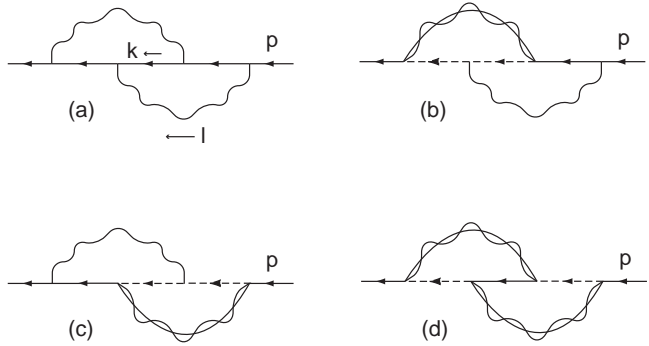}

\caption{Two-loop diagrams with two gauge multiplet propagators.
Solid lines represent electrons, wavy lines represent photons, wavy
lines with a solid core represent photinos, and dashed lines represent
selectrons.}
\label{fig1}

\end{figure}

\begin{figure}[ht]

\epsfxsize 3.35 in \epsfbox{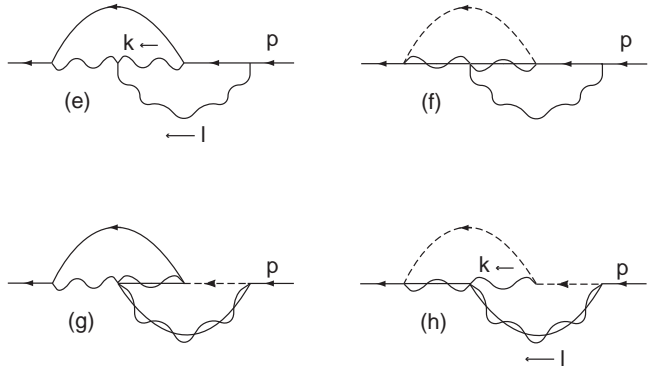}

\caption{Two-loop diagrams with three gauge multiplet propagators.}
\label{fig2}

\end{figure}

Here we isolate the two-loop contribution to the operator in
Eq.~(\ref{eq:danger}).  In ordinary NCQED, there is no one-loop
diagram that contributes to ${\cal O}_1$.  Two-loop diagrams that
contribute are shown in Figs.~\ref{fig1} and~\ref{fig2}.  


We will extract the terms proportional to $\sigma_{\mu\nu}$ and work
on shell ({\em i.e.}, we evaluate the diagrams between spinors $\bar
u(p)$ and $u(p)$ and use $\not \!\! p u(p) = m_e u(p)$.)  For each of
the 8 diagrams, the $\sigma_{\mu\nu}$ terms are proportional to the
electron mass $m_e$.  After extracting the overall electron mass
factor, we set the electron mass and  momentum $p$ to zero in the
integrals as a simplifying assumption.  This leads to corrections in
the final result that are wholly negligible as far as our numerical
analysis is concerned.

Each of the diagrams with only electrons and  photons,  
(a) and (e), give identical results, and the sum of the two is
\begin{equation}
	{\cal M}_a + {\cal M}_e = 24 i m_e e^4 \int (dk) (dl) \,
\frac{ e^{i\, l\cdot\theta\cdot k} \sigma_{\mu\nu} k^\mu l^\nu}
{{k^2 l^4 (k+l)^4}}   
\ ,
\end{equation}
where $(dk) \equiv d^4 k/(2\pi)^4$ and $l\cdot\theta\cdot k 
\equiv l_\mu \theta^{\mu\nu} k_\nu$.  The result (using techniques 
shown below) is 
\begin{equation}   \label{plain}
	{\cal M}_a + {\cal M}_e =  {1 \over 8} m_e \alpha^2
	{\sigma_{\mu\nu} \theta^{\mu\nu} \over 
\sqrt{(-1/2){\rm Tr}\, \theta^2} } \ .
	\end{equation}
	
Now consider the diagrams with superpartners.  The four diagrams with
three superpartner propagators all give the same result, at least to
the operator $\sigma_{\mu\nu} \theta^{\mu\nu}$, and similarly for the
two diagrams with four superpartner propagators. We will give some 
detail of how the diagrams are evaluated.  Using
diagram~(h), as an example, we have
\begin{equation}
	{\cal M}_h = -4 i m_e e^4 \int (dk) (dl) 
\frac{e^{i\,l\cdot\theta\cdot k} \sigma_{\mu\nu} k^\mu l^\nu}{ 
 k^{2} (l^2-M^2)^{2} ((k+l)^2 - M^2)^{2}}  \ ,
	\end{equation}
where $M$ is a common superpartner mass.  We can combine denominators
using a Feynman parameter, and shift one of the integration momenta to
obtain
\begin{equation}
	{\cal M}_h = -24 i m_e e^4 
		\int_0^1 dx \, x (1-x) 
		\int (dk)(dl) \frac{e^{i\,l\cdot\theta\cdot k} 
		\sigma_{\mu\nu} k^\mu l^\nu}{  
	 k^{2} 
		\left[ l^2 + x (1-x) k^2 -M^2 \right]^{4}}  \ .
	\end{equation}
	
Say that only $\theta_{12} = -\theta_{21} \equiv \theta \ne 0$.  Then
\begin{equation}
	{\cal M}_h = 4 m_e e^4 \sigma_{12} 
		{\partial \over \partial \theta} J_h \ ,
	\end{equation}
where after rescaling $k$ and Euclideanizing, we have
\begin{equation}
	J_h = 6 \int_0^1 dx \int (dk)(dl) \,
		\frac{e^{i(l_1 k_2 - l_2 k_1)\theta/\sqrt{x (1-x)} }}{
		 k^{2} \left[l^2 + k^2 +M^2 \right]^{4}}  \ .
	\end{equation}
Now the $dl_0 dl_3$ integrals can be done.  After combining the remaining 
denominators using another Feynman parameter and rescaling the remaining 
components of $l$, we get 
\begin{equation}
	J_h = {3 \over 8\pi^3} \int_0^1 dx \int_0^1 dy \, y
		\int (dk) dl_1 dl_2 \,
		\frac{e^{i(l_1 k_2 - l_2 k_1)\theta/\sqrt{y x (1-x)} }}{
		\left[ k^2 + l_1^2 + l_2^2 + y M^2 \right]^{4}}  
		\ .
	\end{equation}
Now do the $dk_0 dk_3$ integrals, and put the denominators into the 
exponential using a Schwinger parameter.  After one more rescaling of the 
remaining momenta, we have
\begin{eqnarray}
	J_h &=& {1\over 256\pi^6} \int_0^1 dx \int_0^1 dy \, y
		\int_0^\infty dz 
		\int dk_1 dk_2 dl_1 dl_2 \, 
	\nonumber \\    &\times&
		e^{-y z M^2 - k_1^2 - k_2^2 - l_1^2 -l_2^2 
		+ i(l_1 k_2 - l_2 k_1)\theta/
		\big( z \sqrt{y x (1-x)} \big) }
	\nonumber \\
		&=& {1\over 256\pi^4} \int_0^1 dx \int_0^1 dy \, y
		\int_0^\infty dz
	{4 z^2 y x (1-x) \over
		4 z^2 y x (1-x) + \theta^2} 
		e^{-y z M^2} \ .
	\end{eqnarray}
Thus,
\begin{equation}
	{\cal M}_h = -{m_e \alpha^2 \over 2 \pi^2} \sigma_{12}
		\int_0^1 dx \int_0^1 dy \, y \int_0^\infty dz 
		 {4 z^2 y x (1-x) \theta \over
		(4 z^2 y x (1-x) + \theta^2)^2 } 
		e^{-y z M^2} \ .
	\end{equation}
	
The end result for ${\cal M}_f$ is a similar expression, but with the front 
integrals reading 
\begin{equation}
	\int_0^1 dx \, x \int_0^1 dy \, (1-y) \ldots \ .
	\end{equation}
Noting that the rest of the integrand is symmetric under inversion about 
$x = 1/2$, we can replace ``$x$'' in the line above by 
``$(x -1/2) + 1/2$,'' and keep only the ``1/2.''  The 6 graphs involving 
superpartners sum to
\begin{equation}
	4{\cal M}_f + 2{\cal M}_h = -{ m_e \alpha^2 \over  \pi^2} 
		\sigma_{12}
		\int_0^1 dx \int_0^1 dy  \int_0^\infty dz 
		  {4 z^2 y x (1-x) \theta \over
		(4 z^2 y x (1-x) +  \theta^2 )^2} 
		e^{-y z M^2} \ .
	\end{equation}
	
\begin{figure}[h]

\epsfxsize 3.3 in \epsfbox{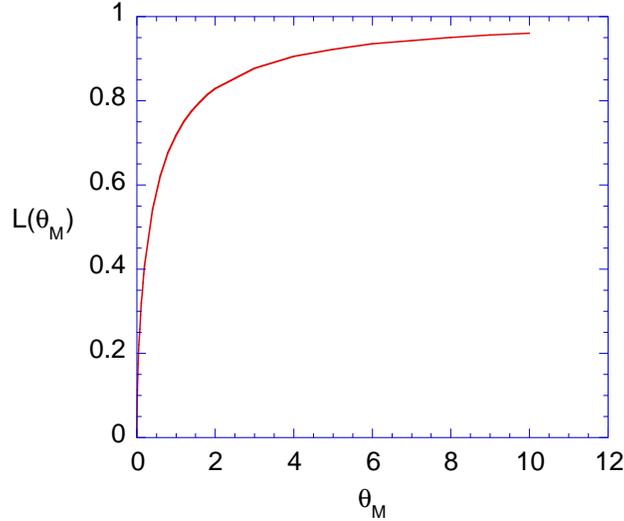}

\caption{The function $L(\theta_M)$.}
\label{lebedfunction}

\end{figure}

In the supersymmetric limit $M \rightarrow 0$, the integrals can be done 
exactly.  For general $M$, it is convenient to rescale $z$,
\begin{equation}
	4{\cal M}_f + 2{\cal M}_h = -{ m_e \alpha^2 \over  \pi^2} 
		{\sigma_{12} \over (\theta M^2)^3}
		\int_0^1 dx \int_0^1 dy  \int_0^\infty dz 
		{4 z^2  x (1-x) \over
		(4 z^2 x (1-x) (\theta M^2)^{-2} +  y )^2} 
		e^{-z} \ .
	\end{equation}
The $y$ and $x$ integrals can both be done, and the full result for the 
two-loop contributions to ${\cal O}_1$ becomes
\begin{equation}          \label{main}
	{\cal M} = \sum_{I=a}^g {\cal M}_I =
		 {1\over 8} m_e \alpha^2 {\sigma_{\mu\nu} \theta^{\mu\nu} 
\over \sqrt{(-1/2){\rm Tr}\, \theta^2} } L(\theta_M)   \ , 
	\end{equation}
where $\theta_M \equiv M^2 \sqrt{(-1/2) {\rm Tr}\, \theta^2}$ and
\begin{equation}
	L(\theta_M) =
		 1 - {2 \theta_M \over \pi^2} 
		\int_0^\infty {dz \over z \sqrt{z^2 + \theta_M^2} }
		e^{-z} \ln{ \sqrt{z^2 + \theta_M^2} + z \over  
					\sqrt{z^2 + \theta_M^2} - z  }  \ .
	\end{equation}
The $z$ integral can be computed analytically, but 
the answer is not enlightening and we do not show it.  Function $L$ 
satisfies $L(0)=0$ and $L(\infty)=1$ and is shown in Fig.~\ref{lebedfunction}.

For the choice $\theta^{0i}=0$~\cite{unitarity}, $\theta^{\mu\nu}$ defines 
a 3-vector in a fixed direction $\hat n$ (where $\hat n$ is a unit vector) 
and the result~(\ref{main}) can be written as a effective Lagrangian
\begin{equation}
{\cal L}_{{\rm eff}} = \frac{1}{2} R_\infty \, L(\theta_M)\, 
\vec\sigma \cdot \hat n
	 \ ,
	\end{equation}
where $R_\infty \equiv m_e \alpha^2/2 = 13.6$ eV.  Searches for such a
term in magnetic systems~\cite{bluhm} show that matrix elements of
${\cal L}_{{\rm eff}}$ are below 10$^{-19}$ eV.  Doing without any cutoff,
$L\rightarrow 1$, is impossible.  One must get a severe suppression
from $L(\theta_M)$, requiring $\theta_M \equiv M^2/\Lambda_{NC}^2
\ll 1$.  The slope of $L$ near the origin is infinite, meaning that
$L(\theta_M)$ has a nonanalytic behavior for $\theta_M \rightarrow 0$.
Numerical evaluations suggest
\begin{equation}
	L(\theta_M) \approx 3\,(\theta_M)^{0.78}
	\end{equation}
for small argument.  From this we estimate $\theta_M \alt 10^{-26}$ or
$\Lambda_{NC} \agt 10^{13} M$.

\section{One-loop Effects}

In this section, we briefly consider Lorentz-violating operators that
are quadratic in the photon field.  We focus on 
\begin{equation}
(\theta_{\alpha\beta}F^{\alpha\beta})^2  \,\,\, ,
\label{eq:photops}
\end{equation}
which is generated at the one-loop level~\cite{ABDG}; here we will 
evaluate the contribution to this operator in the softly-broken 
supersymmetric theory.  

Three one-loop diagrams are relevant to this computation: a photon
loop, a ghost loop, and a photino loop.  Letting $p$ represent the
external photon momentum, and $\ell$ a loop momentum, the relevant
contributions to the photon self-energy from photino, photon and ghost
loops are given respectively by
\begin{equation}
i \Delta\Pi^{\mu\nu} = -16 e^2 I^{\mu\nu}(M) + 20 e^2 I^{\mu\nu}(0)-4 e^2 
I^{\mu\nu}(0)
  \,\,\, ,
\end{equation}
where
\begin{equation}
I^{\mu\nu} = \int (d\ell)
\frac{\ell^\mu\ell^\nu \sin^2[\frac{1}{2}
\ell \cdot \theta\cdot p]}
{[\ell^2-M^2] [(p+\ell)^2-M^2]}  \,\,\, .
\label{eq:olint}
\end{equation}
Again, $M$ represents the photino mass.  In the supersymmetric limit,
$\Delta\Pi^{\mu\nu}$ vanishes.   

In terms of complex exponentials, Eq.~(\ref{eq:olint}) is similar in
form to the integrals discussed in Section 2.  In particular, one may
evaluate the $\mu=\nu=1$ element of $I^{\mu\nu}$, assuming that
$\theta^{0i}=0$~\cite{unitarity}, and that the photon is on-shell and
propagates in the $3$ direction.  The intermediate steps are similar
to those discussed earlier in the two-loop example, so we do not
present them. The correct Lorentz structure of the final result may be
inferred from the specific case.  We find
\begin{equation}
i \Delta \Pi^{\mu\nu} = \frac{i e^2}{8 \pi^2} M^4 \theta^{\mu\alpha} 
p_\alpha \theta^{\nu\beta} p_\beta
I(\xi) \,\,\, ,
\end{equation}
and
\begin{equation}
I(\xi)=\int_0^\infty dt\, \frac{(1-e^{-t})}{t^3} \exp[-\xi/(4t)] 
\,\,\, ,
\end{equation}
where $\xi$ represents the dimensionless combination 
$M^2 (p_\mu \theta^{\mu\alpha} \theta_{\alpha\beta} p^\beta) \equiv
M^2(p \cdot\theta\cdot\theta\cdot p)$. This last integral can be evaluated 
analytically, and expressed in terms of a modified Bessel function:
\begin{equation}
I(\xi) = \frac{16}{\xi^2} \left[1-\frac{\xi}{2}K_2(\sqrt{\xi})\right] \,\,\, .
\end{equation}

To study the phenomenological consequences of this result, it is useful to 
consider the case in which $\xi$ is small (for example, $M$ and $p$ of order
the weak scale with noncommutativity at a high Planck scale).  Notice that 
one may expand the quantity $\xi^2 I(\xi)$ as
\begin{equation}
\xi^2 I(\xi) = 4 \xi + 
\Big(\gamma-\frac{3}{4}-\ln(2)+\frac{1}{2}\ln(\xi)\Big)\xi^2 
+ O(\xi^3)
\,\,\, ,
\end{equation}
from which one may deduce
\begin{equation}
i \Delta \Pi^{\mu\nu} = \frac{i e^2 M^2}{2 \pi^2} 
\left[\frac{\theta^{\mu\alpha}p_\alpha
\theta^{\nu\beta} p_\beta}{p\cdot\theta\cdot\theta\cdot p}\right] + 
\frac{i e^2 M^4}{8 \pi^2} \Big(\gamma-\frac{3}{4}-\ln(2)\Big) \left[
\theta^{\mu\alpha}p_\alpha \theta^{\nu\beta} p_\beta\right] + \cdots \,\,\, .
\label{eq:expdform}
\end{equation}
The result for off-shell photons may be obtained by replacing $M^2$ by
$M^2-z(1-z)p^2$ and integrating the final result between $z=0$ and
$1$. The second term in the expansion~(\ref{eq:expdform}) contributes 
directly to the operator of interest in Eq.~(\ref{eq:photops}).  
The dependence on $M$ is the same as one would expect in a theory 
regulated by a momentum-space cutoff, and we therefore obtain the same bound
given in Ref.~\cite{ABDG}, from limits on birefringent effects in light
from distant galaxies, $\theta M^2< 10^{-12}$.  This is much weaker than
the bound obtained in Section~2, $\theta M^2 \alt 10^{-26}$. The first 
term in Eq.~(\ref{eq:expdform}) is peculiar in that it is independent of 
the scale of $\theta$.  In Ref.~\cite{lenny}, it was suggested that such 
terms vanish in the case of softly-broken supersymmetry, in conflict with 
our explicit result.  This term leads to a tachyonic photon~\cite{armoni}
for certain polarizations, with a mass scale $\sim
\sqrt{\alpha/\pi} \, M \sim 100$~GeV for $M\sim 1$~TeV, independent
of the scale of $\theta$.  If this term is physical, it rules out
canonical NCQED by itself, unless there is a further modification of the 
theory in the ultraviolet.  However, it has also been suggested that terms 
like the first in Eq.~(\ref{eq:expdform}) may be artifacts related to the 
use of Wess-Zumino gauge~\cite{grim}.  We therefore consider our result 
from Section~2 as a more conservative bound on the scale of noncommutativity.


\section{Conclusions}


We have studied Lorentz-violating operators in softly-broken,
supersymmetric noncommutative QED.  In the limit where the
superpartners decouple, we recover the nonsupersymmetric result that
the most dangerous operators lead to conflict with experimental
bounds~\cite{ABDG}.  In the opposite limit where supersymmetry is
unbroken, the operators of interest are forbidden exactly.  This
statement follows because there is no supersymmetric way to write
these operators; alternatively, one may see that the Feynman diagrams
that contribute to the most dangerous operators cancel exactly.
Including a soft mass $M$ for the superpartners leads to non-zero
values for the operator coefficients; the supersymmetry-breaking mass
thus acts as a gauge-invariant regulator, allowing one to interpolate
between these limits. In contrast to the results obtained by applying
a simple momentum-space cutoff, the dependence on the superpartner
mass $M$ is not analytic as $M$ goes to zero.  The bound that follows
from searches for Lorentz violation in magnetic systems~\cite{bluhm}
is seven orders of magnitude more severe, $\theta M^2 \alt 10^{-26}$,
and places the scale of noncommutativity at or above the conventional
supersymmetric grand unification scale for $M\sim 1$~TeV.  

If nature uses noncommutative coordinates, it need not be done with a
Lorentz-violating implementation.  One may take the results of the
present investigation as motivation to pursue space-time
noncommutativity in Lorentz-covariant ways~\cite{ncxd,nclc}.


\appendix
\section{Noncommutative Feynman Rules}

\begin{figure}[h]

\begin{tabular}{lc}
	\epsfxsize 1 in \epsfbox{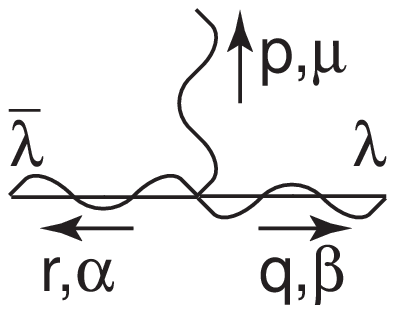}  
&	\raisebox{4.5ex}{  \quad
		$2e \sin({1\over 2} r \cdot \theta \cdot q)
		( \gamma_\mu)_{\alpha\beta}$
					}
\\
	\epsfxsize 1 in \epsfbox{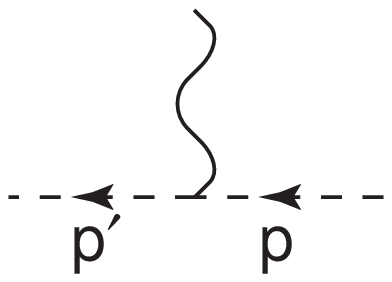}  
&	\raisebox{4.5ex}{  \quad
		$-ie \exp({i\over 2} p \cdot \theta \cdot p')
		(p'+p)_\mu$
					}
\\
	\epsfxsize 1 in \epsfbox{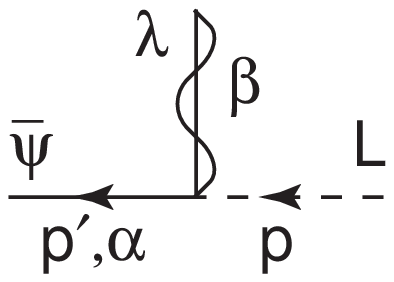}  
&	\raisebox{4.5ex}{  \quad
		${ \displaystyle{e} \over\sqrt{2} } 
		( 1 + \gamma_5 )_{\alpha\beta}
		\exp({i\over 2} p \cdot \theta \cdot p')$
					}
\\
	\epsfxsize 1 in \epsfbox{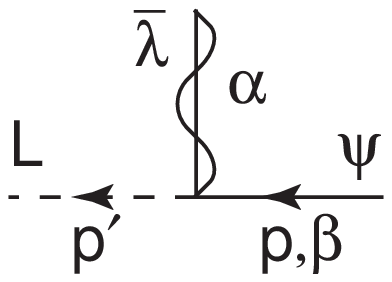}  
&	\raisebox{4.5ex}{  \quad
		$ {-\displaystyle{e} \over\sqrt{2} } 
		( 1 - \gamma_5 )_{\alpha\beta}
		\exp({i\over 2} p \cdot \theta \cdot p')$
					}
\end{tabular}

\caption{Feynman rules for superpartners in noncommutative
supersymmetric QED.  The rules for the right-handed scalars can be
obtained from the left-handed ones shown by $\gamma_5 \rightarrow
-\gamma_5$. $\psi$ and $\lambda$ represent electrons and photinos,
respectively.  Our sign conventions are based on those of Ref.~\cite{BL}.}

\label{feynman}

\end{figure}

The superpartner noncommutative QED Feynman rules that are needed for
this paper are given in Fig.~\ref{feynman}.  The remaining Feynman rules can
be found in the literature (e.g.,~\cite{lenny}).


%
\begin{acknowledgments}
C.E.C. and C.D.C. thank the NSF for 
support under Grant No.\ PHY-9900657.  C.D.C. thanks the Jeffress 
Memorial Trust for support under No.\ Grant J-532. R.F.L. thanks the 
DOE for support under Grant No.\ DE-AC05-84ER40150 and 
the NSF for support under No.\ Grant PHY-0140362.  We thank A.~Armoni and
L.~Susskind for useful comments.
\end{acknowledgments}


\begin{thebibliography}{99}

\bibitem{ncqed1}
J.~L.~Hewett, F.~J.~Petriello and T.~G.~Rizzo, 
Phys.\ Rev.\ D {\bf 64}, 075012 (2001);  
S.~Godfrey and M.~A.~Doncheski, 
Phys.\ Rev.\ D {\bf 65}, 015005 (2002); 
N.~Mahajan, 
hep-ph/0110148;
S.~w.~Baek, D.~K.~Ghosh, X.~G.~He and W.~Y.~Hwang,
Phys.\ Rev.\ D {\bf 64}, 056001 (2001);
H.~Grosse and Y.~Liao,
Phys.\ Rev.\ D {\bf 64}, 115007 (2001);
Phys.\ Lett.\ B {\bf 520}, 63 (2001).

\bibitem{ncsm}
I.~Hinchliffe and N.~Kersting, 
Phys.\ Rev.\ D {\bf 64}, 116007 (2001) and
hep-ph/0205040;
E.~O.~Iltan,
Phys.\ Rev.\ D {\bf 66}, 034011 (2002) and
New J.\ Phys.\  {\bf 4}, 54 (2002);
H.~Falomir {\em et al.},
Phys.\ Rev.\ D {\bf 66}, 045018 (2002);
S.~Narison,
hep-ph/0208225;
J.~i.~Kamoshita,
hep-ph/0206223.

\bibitem{CHKLT}
S.~M.~Carroll {\em et al.},
Phys.\ Rev.\ Lett.\  {\bf 87}, 141601 (2001).

\bibitem{MPR}
I.~Mocioiu, M.~Pospelov and R.~Roiban,
Phys.\ Lett.\ B {\bf 489}, 390 (2000).

\bibitem{ABDG}
A.~Anisimov, T.~Banks, M.~Dine and M.~Graesser,
Phys.\ Rev.\ D {\bf 65}, 085032 (2002).

\bibitem{CCL}
C.~E.~Carlson, C.~D.~Carone and R.~F.~Lebed,
Phys.\ Lett.\ B {\bf 518}, 201 (2001).

\bibitem{ncxd}
C.~E.~Carlson and C.~D.~Carone,
Phys.\ Rev.\ D {\bf 65}, 075007 (2002).

\bibitem{GMW}
J.~Gomis, T.~Mehen and M.~B.~Wise,
JHEP {\bf 0008}, 029 (2000).

\bibitem{posp2}
I.~Mocioiu, M.~Pospelov and R.~Roiban,
Phys.\ Rev.\ D {\bf 65}, 107702 (2002).

\bibitem{SW}
N.~Seiberg and E.~Witten,
JHEP {\bf 9909}, 032 (1999).

\bibitem{dn}
See, for example,
M.~R.~Douglas and N.~A.~Nekrasov,
Rev.\ Mod.\ Phys.\  {\bf 73}, 977 (2001).

\bibitem{CK}
D.~Colladay and V.~A.~Kostelecky,
Phys.\ Rev.\ D {\bf 58}, 116002 (1998);
See also, V.~A.~Kostelecky {\em et al.},
Phys.\ Rev.\ D {\bf 65}, 056006 (2002)
and hep-ph/0205211.

\bibitem{ncs}
S.~Terashima,
Phys.\ Lett.\ B {\bf 482}, 276 (2000);
C.~S.~Chu {\em et al.},
Phys.\ Lett.\ B {\bf 513}, 200 (2001).

\bibitem{unitarity}
J.~Gomis and T.~Mehen,
Nucl.\ Phys.\ B {\bf 591}, 265 (2000).

\bibitem{bluhm}
R.~Bluhm and V.~A.~Kostelecky,
Phys.\ Rev.\ Lett.\  {\bf 84}, 1381 (2000).


\bibitem{lenny}
A.~Matusis, L.~Susskind and N.~Toumbas,
JHEP {\bf 0012}, 002 (2000).

\bibitem{armoni}
A.~Armoni and E.~Lopez,
Nucl.\ Phys.\ B {\bf 632}, 240 (2002)

\bibitem{grim}
A.~A.~Bichl {\it et al.},
hep-th/0203141.

\bibitem{nclc}
C.~E.~Carlson, C.~D.~Carone and N.~Zobin,
hep-th/0206035 (to appear in Phys.\ Rev.\ D).

\bibitem{BL}
D.~Bailin and A.~Love,
{\it Supersymmetric Gauge Field Theory And String Theory},
Institute of Physics Publishing, Bristol, UK, 1994.

\end{thebibliography}

\end{document}